\documentstyle[lscape]{mn}

\begin{document}

\title[Estimating photometric redshifts with ANNs]{Estimating photometric redshifts with artificial neural networks}
\author[A. E. Firth, O. Lahav, R. S. Somerville]
       {Andrew E. Firth$^{1}$, Ofer Lahav$^{1}$\thanks{lahav@ast.cam.ac.uk}, Rachel S. Somerville$^{1,2}$\\
       $^{1}$ Institute of Astronomy, University of Cambridge, Cambridge, CB3 0HA, UK\\
       $^{2}$ Department of Astronomy, University of Michigan, Ann Arbor, MI48109--1090, USA\\}
\date{Submitted to MNRAS, 15 March 2002}

\pagerange{\pageref{firstpage}--\pageref{lastpage}}
\pubyear{20??}

\maketitle

\label{firstpage}

\begin{abstract}
A new approach to estimating photometric redshifts -- using Artificial Neural Networks (ANNs) -- is investigated.  Unlike the standard template-fitting photometric redshift technique, a large spectroscopically-identified training set is required but, where one is available, ANNs produce photometric redshift accuracies at least as good as and often better than the template-fitting method.  The Bayesian priors on the underlying redshift distribution are automatically taken into account.  Furthermore, inputs other than galaxy colours -- such as morphology, angular size and surface brightness -- may be easily incorporated, and their utility assessed.

Different ANN architectures are tested on a semi-analytic model galaxy catalogue and the results are compared with the template-fitting method.  Finally the method is tested on a sample of $\sim$20000 galaxies from the Sloan Digital Sky Survey.  The r.m.s.\ redshift error in the range $z \la 0.35$ is $\sigma_z \sim 0.021$.
\end{abstract}


\section[]{Introduction} \label{sec.intro}
The basic photometric redshift technique is to use the colours of a galaxy in a selection of medium- or broad-band filters as a crude approximation of the galaxy's spectral energy distribution or SED, in order to find its redshift and spectral type.  The technique is very efficient compared with spectroscopic redshifts since the signal-to-noise in broad-band filters is much greater than the signal-to-noise in a dispersed spectrum and, furthermore, a whole field of galaxies may be imaged at once while spectroscopy is limited to individual galaxies or those that can be positioned on slits or fibres.  However photometric redshifts are only approximate at best and are sometimes subject to complete misidentifications.  For many applications though, large sample sizes are more important than precise redshifts and photometric redshifts may be used to good effect.

Photometric redshifts date back to Baum (1962; see also Hogg et al.\ 1998; Weymann et al.\ 1999).  They have been used extensively in recent years on the ultra-deep and well-calibrated Hubble Deep Field observations (e.g.\ Gwyn \& Hartwick 1996; Connolly, Szalay \& Brunner 1998; Fern\'andez-Soto, Lanzetta \& Yahil 1999; Fontana et al.\ 2000; Fern\'andez-Soto et al.\ 2001; Massarotti, Iovino \& Buzzoni 2001a; Massarotti et al.\ 2001b).  The most commonly used approach is the template-fitting technique.  This involves compiling a library of template spectra -- either theoretical SEDs from population synthesis models (e.g.\ {\tt GISSEL} -- Bruzual \& Charlot 1993) or empirical SEDs (e.g.\ Coleman, Wu \& Weedman 1980).  Then the expected flux through each survey filter is calculated for each template SED on a grid of redshifts, with corrections for ISM, IGM and Galactic extinction where necessary.  A redshift and spectral type are estimated for each observed galaxy by minimizing $\chi^2$ with respect to redshift, $z$, and spectral type, SED, where
\begin{equation}
\chi^2(z,{\mathrm{SED}})=\sum_i\left(\frac{f_i-\alpha(z,{\mathrm{SED}})t_i(z,{\mathrm{SED}})}{\sigma_i}\right)^2, \label{eq.photz.chi2}
\end{equation}
$f_i$ is the observed flux in filter $i$, $\sigma_i$ is the error in $f_i$, $t_i(z,{\mathrm{SED}})$ is the flux in filter $i$ for the template SED at redshift $z$ and $\alpha(z,{\mathrm{SED}})$ (the scaling factor normalizing the template to the observed flux) is determined by minimizing equation \ref{eq.photz.chi2} with respect to $\alpha$, giving
\begin{equation}
\alpha(z,{\mathrm{SED}})=\left(\sum_i\frac{f_it_i(z,{\mathrm{SED}})}{\sigma_i^2}\right)/\left(\sum_i\frac{t_i(z,{\mathrm{SED}})^2}{\sigma_i^2}\right).
\end{equation}

The template-fitting photometric redshift technique makes use of the available and reasonably detailed knowledge of galaxy SEDs and in principle it may be used reliably even for populations of galaxies for which there are few or no spectroscopically confirmed redshifts.  However, crucial to its success, is the compilation of a library of accurate and representative template SEDs (see e.g.\ Hogg et al.\ 1998; Firth 2002b).  Empirical templates are typically derived from nearby bright galaxies, which may not be truly representative of high redshift galaxies.  Conversely, while theoretical SEDs can cover a large range of star formation histories, metallicities, dust extinction models etc., not all combinations of these parameters (at any particular redshift) are realistic, and the ad hoc inclusion of superfluous templates increases the potential for misidentifications when using observations with noisy photometry.

An alternative approach can be used when one has a sufficiently large (e.g.\ $\sim$100--1000, depending on the redshift range) and representative subsample with spectroscopic redshifts.  Then one can fit a polynomial or other function mapping the photometric data to the known redshifts and use this to estimate redshifts for the remainder of the sample with unknown redshifts (e.g.\ Connolly et al.\ 1995b; Brunner, Szalay \& Connolly 2000; Sowards-Emmerd et al.\ 2000).  With this approach, errors in the estimated redshifts may also be estimated analytically or via Monte Carlo simulations.

An extension of the latter approach is to use Artificial Neural Networks (ANNs hereafter).  ANNs have been used before in astronomy for, amongst other things, galaxy morphological classification (e.g.\ Storrie-Lombardi et al.\ 1992; Naim et al.\ 1995; Lahav et al.\ 1996), morphological star/galaxy separation (e.g.\ Bertin \& Arnouts 1996; Andreon et al.\ 2000) and stellar spectral classification (e.g.\ Bailer-Jones, Irwin \& von Hippel 1998; Allende Prieto et al.\ 2000; Weaver 2000).  Essentially an ANN takes a set of inputs (e.g.\ logarithms of fluxes -- i.e.\ magnitudes -- in different filters) for each object, applies some non-linear function, and outputs a value (e.g.\ the estimated redshift).  The ANN is first trained -- i.e.\ the coefficients (weights) of the function are optimized -- by using a training set where the desired output is known.  The ANN may then be used on any number of other objects with similar inputs (i.e.\ magnitudes in the same filter set) but unknown outputs (i.e.\ redshifts).

As well as using all of the information contained in the magnitudes and colours, provided the training set is a representative subsample of the data, the ANN will also take into account the Bayesian priors on the galaxy redshift distribution (cf.\ Benitez 1998; Teplitz et al.\ 2001).  While choosing a template library that is both sufficient and non-superfluous is a source of concern for the template-fitting method, ANNs automatically fit the true range of galaxy SEDs.  Another potential advantage of ANNs relative to the template-fitting method is that the weights applied to each filter may be more optimal than simple $\chi^2$-weighting.  In addition one can also feed in other observational input such as image size or surface brightness, morphology and concentration parameters where such data are available.  It is interesting then to see how the two methods compare.

This paper explores the use of ANNs as a potential tool for photometric redshift determination.  The layout of this paper is as follows.  In $\S$\ref{sec.anns} the ANNs are described and in $\S$\ref{sec.sam} a semi-analytic model (used to provide a simulated galaxy catalogue) is introduced.  In $\S$\ref{sec.cf} the ANN parameters (architecture and training set size) are investigated using the simulated galaxy catalogue and in $\S$\ref{sec.hyperz} the performance of ANNs are compared with the performance of the traditional template-fitting method.   $\S$\ref{sec.hist} looks at the effect of photometric noise and in $\S$\ref{sec.sed} ANNs are investigated as a method for also determining spectral type from redshifted data.  In $\S$\ref{sec.sdss}, ANNs are tested on Sloan Digital Sky Survey observational data.  The science prospects are briefly discussed in $\S$\ref{sec.conclude}.

\section[]{Artificial neural networks}  \label{sec.anns}

\begin{figure}
\vspace{6.8cm}
\includegraphics{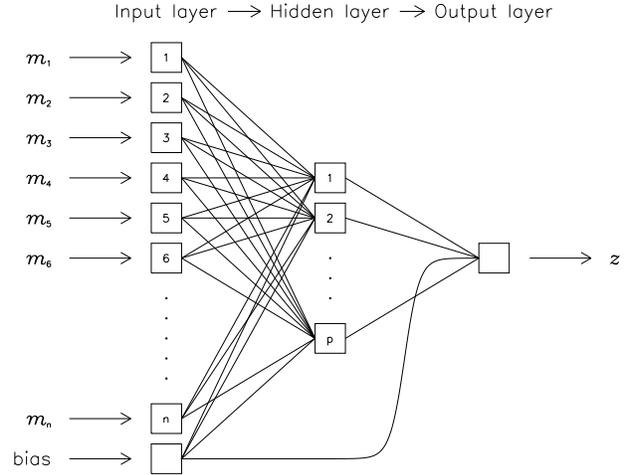}
\caption{\label{fig.photz.arch} A schematic diagram of an ANN with input nodes taking, for example, magnitudes $m_i = -2.5\log_{10}f_i$ in various filters, a single hidden layer, and a single output node giving, for example, redshift $z$.  The architecture is $n$:$p$:1 in the notation used in this paper.  Each connecting line carries a weight $w_j$. The bias node allows for an additive constant in the network function defined at each node.  More complex nets can have additional hidden layers.}
\end{figure}

An ANN comprises a set of input nodes, one or more output nodes, and one or more hidden layers each containing a number of nodes (Fig.\ \ref{fig.photz.arch}; see e.g.\ Bishop 1995, and references therein, for background).  A particular network architecture may be denoted by $N_{\mathrm{in}}$:$N_1$:$N_2$:...:$N_{\mathrm{out}}$ where $N_{\mathrm{in}}$ is the number of input nodes, $N_1$ is the number of nodes in the first hidden layer, and so on.  For example 9:6:1 takes 9 inputs, has 6 nodes in a single hidden layer and gives a single output.  The nodes are connected and each connection carries a weight which together comprise the vector of coefficients $\bmath{w}$ which are to be optimized.  Unless otherwise stated, here every node is assumed to be connected to every node in the previous layer and to every node in the next layer only, but it is certainly possible to have more or less interconnected nets.  The input parameters for each object are represented by the vector $\bmath{x}$ (e.g.\ the magnitudes in a set of filters). Given a training set of inputs $\bmath{x}_k$ and desired outputs $z_k$ (e.g.\ the redshift), the ANN is optimized by minimizing the cost function
\begin{equation}
E=\frac{1}{2}\sum_k[z_k-F(\bmath{w},\bmath{x}_k)]^2.  \label{eq.ann.e}
\end{equation}
The function $F(\bmath{w},\bmath{x}_k)$ is given by the network.  A function $g_p$ is defined at each node $p$, taking as its argument
\begin{equation}
u_p=\sum_j w_j x_j  
\end{equation}
where the sum is over the input nodes to $p$.  These functions are typically taken (in analogy to biological neurons) to be sigmoid functions such as $g_p(u_p)=1/[1+\exp(-u_p)]$ (used here).  An extra input node -- the bias node -- is automatically included to allow for additive constants in these functions.  The combination of these functions over all the network nodes makes up the function $F$.  A programme kindly provided by B. D. Ripley was used to train the networks.  The programme takes as its input a network architecture, a training set and a random seed to initiate the weight vector $\bmath{w}$\footnotemark\footnotetext{The initial weights were randomly chosen from a uniform distribution with range [$-$0.7, 0.7].}, and uses an iterative quasi-Newton method (see e.g.\ Bishop 1995) to minimize the cost function.  To ensure that the weights are regularized (i.e.\ that they do not become too large), an extra quadratic cost term
\begin{equation}
E_w=\beta\frac{1}{2}\sum_j w_j^2, \label{eq.ann.ew}
\end{equation}
was added to equation \ref{eq.ann.e}.  A value of $\beta = 0.0001$ was chosen empirically to optimize the ANN performance.  After each training iteration, the cost function is evaluated on a separate validation set.  After a chosen number of training iterations, training terminates and the final weights chosen for the ANN are those from the iteration at which the cost function is minimal on the validation set.  This is useful to avoid over-fitting to the training set if the training set is small.

\section[]{Model galaxy catalogues} \label{sec.sam}
\subsection[]{Semi-analytic models}
To provide a galaxy catalogue on which to train and test ANNs, a semi-analytic model was used.  Semi-analytic models are an attempt to use simple recipes to parameterize the main physical processes of galaxy formation within the hierarchical paradigm of galaxy formation (e.g.\ Kauffmann, White \& Guiderdoni 1993; Cole et al.\ 1994).  In these models, Monte Carlo techniques may be used to efficiently generate large mock galaxy catalogues with a (broadly-speaking) realistic distribution of galaxy types, luminosities, colours and redshifts.  Here the current version of the code developed by Somerville (1997) is used. This has been shown to produce good agreement with many properties of local and high-redshift galaxies (Somerville \& Primack 1999; Somerville, Primack \& Faber 2001; Firth et al.\ 2002a).

In this model (see Somerville \& Primack 1999 for details), the number density of haloes of various masses at a given redshift is determined by an improved version of the Press-Schechter model (Sheth \& Tormen 1999) and the formation and merging of dark matter haloes as a function of time is represented by a `merger tree'.  The cooling of gas, formation of stars, and reheating and ejection of gas by supernovae within these haloes are modelled by simple recipes.  Cold gas is assumed to initially cool into, and form stars within, a rotationally supported disc.  Major mergers between galaxies destroy the discs and create spheroids.  Galaxy mergers also produce bursts of star formation.  The chemical evolution and star formation history of each galaxy is traced and convolved with multi-metallicity stellar population synthesis models (Devriendt, Guiderdoni \& Sadat 1999), and a dust extinction law, in order to calculate the galaxy's SED.

There are several advantages in using a semi-analytic model here.  Firstly, an arbitrarily large number of galaxies may be generated over any desired redshift range, and with any magnitude limit.  The `true' redshift and magnitudes (prior to the addition of photometric noise) are known precisely.  At present there is no large observed spectroscopic sample at high redshift, and those spectroscopic samples that do exist tend to be biased towards luminous galaxies with prominent emission lines at optical wavelengths.  On the other hand, simpler model galaxy catalogues (e.g.\ PLE models) are less likely to produce realistic distributions of galaxy SEDs in terms of composite stellar populations, ages, metallicities and the effects of dust (all as a function of redshift).  Where (when) suitable spectroscopic samples exist, the model catalogues could be replaced by observed photometric and spectroscopic samples and the results would be expected to be comparable (see e.g.\ $\S$\ref{sec.sdss}).

\subsection[]{Preparing the input catalogue}
An $H<22$ catalogue in $UBVRIH$ was generated using the semi-analytic model.  To simulate a real galaxy survey, photometric noise was added to this catalogue, simulating 5$\sigma$ magnitude limits of $U=25.1$, $B=26.6$, $V=26.1$, $R=25.6$, $I=24.7$ and $H=20.5$ (typical of current and future large surveys aimed at studying large-scale structure at high redshifts -- e.g.\ the LCIR Survey, Firth et al.\ 2002a).  An extra 0.05 mags r.m.s.\ error term was included to simulate seeing variations, zeropoint inaccuracies and other sources of photometric errors.  Finally an $H<20.5$ `noisy' sample was drawn from this catalogue.  

So that all weights are treated fairly equally in equation \ref{eq.ann.ew}, it is useful to normalize the magnitudes in each filter and the model redshifts to the range \mbox{[0, 1]}.  The exact form of the normalization is unimportant provided the same normalization is used for both training the ANN and using the ANN.  For definiteness, in each filter the mean magnitude (derived from an $H<20.5$ `noiseless' sample) was subtracted, and the range \mbox{[$-$5, 5]} was mapped linearly to \mbox{[0, 1]}.  Furthermore the redshift range \mbox{[0, 3.5]} was mapped linearly to \mbox{[0, 1]}.

\section{Selecting network parameters} \label{sec.cf}

\subsection{Number of training iterations}
\begin{figure}
\vspace{9.1cm}
\includegraphics{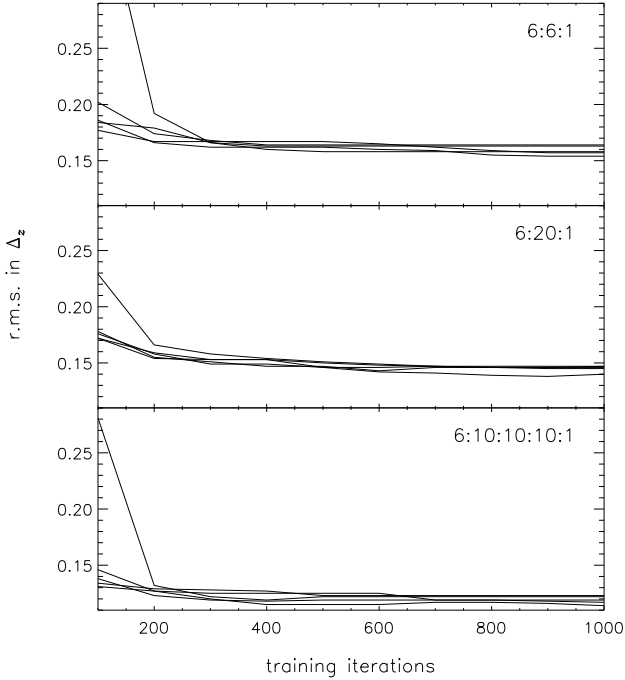}
\caption{\label{fig.photz.anns_niter} The r.m.s.\ in $\Delta_z = z_{\mathrm{model}} - z_{\mathrm{phot}}$ (measured on 10000 test galaxies) as a function of the number of training iterations for three ANN architectures (indicated in upper right of each panel).  For each architecture, five ANNs were trained, each initialized with a different random seed.  A training set of size 10000 was used.}
\end{figure}

First of all, the required number of training iterations $N_{\mathrm{iter}}$ was investigated.  Clearly this will depend on the characteristics of the data set and the complexity of the network architecture.  There is also an element of chance due to the randomized initial weights.  After an ANN has been trained, its performance is assessed by running a {\it testing} set (distinct from the {\it training} and {\it validation} sets) through it and calculating the r.m.s.\ in $\Delta_z = z_{\mathrm{model}} - z_{\mathrm{phot}}$, where $z_{\mathrm{model}}$ are the model redshifts and $z_{\mathrm{phot}}$ are the corresponding ANN photometric redshifts.  Fig.\ \ref{fig.photz.anns_niter} plots the change in r.m.s.\ as $N_{\mathrm{iter}}$ is increased.  Three architectures, covering a range in complexity, are compared and, for each architecture, five ANNs were produced starting with different random seeds.  A training set of size 10000 was used (cf.\ $\S$\ref{sec.cf}.3) and the ANNs were tested on a separate testing set, also of size 10000.  Note that the testing set could be any size.  A sample of size 10000 was chosen simply to provide good statistics.

The r.m.s.\ decreases as $N_{\mathrm{iter}}$ is increased but levels off for large $N_{\mathrm{iter}}$.  For most random seeds, improvement beyond $N_{\mathrm{iter}} \sim 500$ is slow.  For the remainder of this paper, in which a similar range of architectures, and the same -- or simpler -- data sets are considered, $N_{\mathrm{iter}}$ will be restricted to 1000.

\begin{figure}
\vspace{5.5cm}
\includegraphics{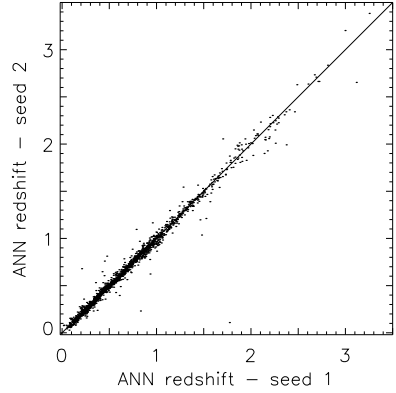}
\caption{\label{fig.photz.anns_seed} Comparison of photometric redshifts (using 2000 test SAM galaxies) for two ANNs initialized with different random seeds.  A 6:10:10:10:1 network architecture, 10000 training galaxies and 1000 training iterations were used.  The r.m.s.\ in $\Delta_z = z_{\mathrm{1}} - z_{\mathrm{2}}$ is 0.065.}
\end{figure}

Fig.\ \ref{fig.photz.anns_seed} compares the estimated photometric redshifts on the testing set, using the 6:10:10:10:1 architecture, for two different random seeds after 1000 training iterations.  The two ANNs closely agree.  Generally, different initial random seeds lead to ANNs with similar r.m.s.\ accuracy, though there are still differences at the $\la$10 per cent level.  It is useful, therefore, to generate several ANNs using different random seeds.  One may then use the validation set to choose the `best' ANN.  A better approach (see e.g.\ Bishop 1995 for details) is to combine a set of ANNs generated using different random seeds.  This is called a `committee of ANNs'.  In this paper, the estimated redshift for each galaxy presented to the committee is taken to be the median of the estimates provided by the individual ANNs.  Typically the committee gives more accurate redshift estimates than any of its component ANNs taken individually (see $\S$\ref{sec.cf}.2, Table \ref{table.photz.anns_arch} for examples).

\subsection{Network architecture}
More complex network architectures have more free parameters (weights) and therefore allow a closer fit to the data.  In any real data set there will be a fundamental limit to the r.m.s.\ fit, due to random noise in the input measurements, and further increases in architecture complexity will provide no significant improvement.  In addition, architectures with more weights take longer to train.  In any given situation one would like to use the simplest network possible while still obtaining optimal results.

\begin{figure}
\vspace{6.4cm}
\includegraphics{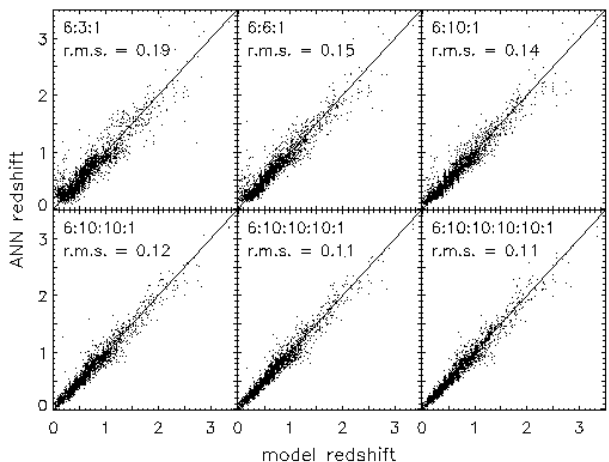}
\caption{\label{fig.photz.anns_arch} Photometric redshift versus model redshift comparisons (using 2000 test galaxies) for several network architectures (indicated at upper left in each panel).  The simulated SAM catalogue is $H<20.5$ with 5$\sigma$ limits $U=25.1$, $B=26.6$, $V=26.1$, $R=25.6$, $I=24.7$ and $H=20.5$.  A committee of 5 ANNs (for each architecture) and a training set of 10000 galaxies were used.  Note the increased scatter at high-redshift where there are fewer training galaxies.  The quantization at 0.1 intervals in $z$ is due to poor redshift-space resolution in the colour grid used to determine galaxy colours in the semi-analytic model.}
\end{figure}

Table \ref{table.photz.anns_arch} compares the r.m.s.\ in $\Delta_z = z_{\mathrm{model}} - z_{\mathrm{phot}}$, evaluated on a testing set of size 10000, for several architectures (see also Fig.\ \ref{fig.photz.anns_arch}).  For committees of five ANNs, the networks with a single hidden layer (i.e.\ 6:$m$:1) reach a limiting r.m.s.\ of $\sim$0.14 (e.g.\ for $m \sim 15$).  There is no significant improvement for larger $m$.  However adding extra hidden layers offers some improvement, even when the total number of weights is not increased.  The committees of five 6:6:6:6:1, 6:10:10:1 and 6:10:10:10:1 ANNs in Table \ref{table.photz.anns_arch} all produce r.m.s.\ values of $\sim$0.115.  The scatter in the r.m.s.\ between different committees is of order $\pm 0.002$ for these architectures, and increasing the number of ANNs in the committee leads only to a small improvement in the r.m.s.\ (e.g.\ $\sim$0.001 decrease in the r.m.s.\ for a committee of 25 ANNs).  Adding further hidden layers (e.g.\ 6:10:10:10:10:1) leads to little or no further improvement.   

For the remainder of this paper, a 6:6:6:6:1 architecture will be used as the fiducial ANN, since this architecture gives results comparable with the best results in Table \ref{table.photz.anns_arch} but has relatively few weights.  Unless stated otherwise, redshifts will be estimated using a committee of five such ANNs -- taking the median redshift estimate of five for each galaxy.

Clearly the required network complexity depends on the data set.  For data with a higher signal-to-noise ratio, the fundamental r.m.s.\ limit on $\Delta_z$ will be lower and a more complex architecture may be necessary to reach this limit.  Conversely, for data covering a smaller redshift range, fewer free parameters will be necessary to model the mapping from colours to redshift, and a simpler architecture may suffice.

\begin{table*}
\caption{\label{table.photz.anns_arch} The mean and r.m.s.\ in $\Delta_z = z_{\mathrm{model}} - z_{\mathrm{phot}}$ for various network architectures.  The simulated SAM catalogue is $H<20.5$ with 5$\sigma$ limits $U=25.1$, $B=26.6$, $V=26.1$, $R=25.6$, $I=24.7$ and $H=20.5$.  A training set of 10000 galaxies was used.  Columns 3 and 4 display mean values for five ANNs initialized with different random seeds.  Columns 5 and 6 display values for a single committee comprising five ANNs (initialized with different random seeds).  Note that a committee produces significantly better results than its component ANNs.}
\begin{center}
\begin{tabular}{lrrrrr}
network     &&\multicolumn{2}{c}{individual ANN}&\multicolumn{2}{c}{committee}\\
architecture&$N_{\mathrm{weights}}$&mean $\Delta_z$&r.m.s.\ in $\Delta_z$&mean $\Delta_z$&r.m.s.\ in $\Delta_z$\\
\hline
6:3:1          &  25& $-$0.005& 0.192& $-$0.004& 0.187\\
6:6:1          &  49& $-$0.005& 0.159& $-$0.004& 0.151\\
6:10:1         &  81& $-$0.004& 0.148& $-$0.005& 0.142\\
6:15:1         & 121& $-$0.003& 0.146& $-$0.004& 0.140\\
6:20:1         & 161& $-$0.004& 0.145& $-$0.004& 0.140\\
6:30:1         & 241& $-$0.003& 0.140& $-$0.003& 0.136\\
6:40:1         & 321& $-$0.003& 0.143& $-$0.003& 0.139\\
6:6:6:1        &  91& $-$0.002& 0.133& $-$0.002& 0.124\\
6:6:6:6:1      & 133& $-$0.001& 0.123& $-$0.000& 0.116\\
6:10:10:1      & 191& $-$0.001& 0.124& $-$0.000& 0.116\\
6:10:10:10:1   & 301& $-$0.000& 0.119&    0.000& 0.113\\
6:10:10:10:10:1& 411&    0.000& 0.119&    0.000& 0.113\\
\hline
\end{tabular}
\end{center}
\end{table*}

\subsection{Size of training set}
\begin{figure}
\vspace{5.cm}
\includegraphics{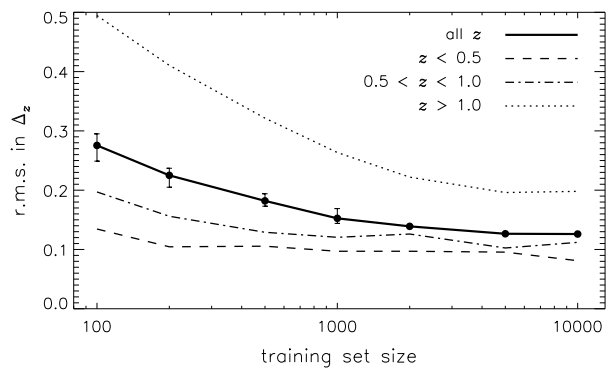}
\caption{\label{fig.photz.anns_ntrain} Photometric redshift accuracy as a function of training set size $N_{\mathrm{train}}$.  The plot shows the r.m.s.\ in $\Delta_z = z_{\mathrm{model}} - z_{\mathrm{phot}}$, evaluated on 10000 test galaxies.  The simulated SAM catalogue is $H<20.5$ with 5$\sigma$ limits $U=25.1$, $B=26.6$, $V=26.1$, $R=25.6$, $I=24.7$ and $H=20.5$.  A 6:6:6:6:1 network architecture (with a single random seed) was used.  For $N_{\mathrm{train}} < 10000$, up to 10 ANNs were generated using separate training sets; points and error bars show respectively medians and interquartile ranges for the different ANNs.}
\end{figure}

Often one will have no choice concerning the size of the training set, $N_{\mathrm{train}}$.  However, when designing surveys, it is useful to assess what size of training set is necessary to provide a given redshift accuracy.  Fig.\ \ref{fig.photz.anns_ntrain} plots the r.m.s.\ in $\Delta_z = z_{\mathrm{model}} - z_{\mathrm{phot}}$, evaluated on a testing set of size 10000, as a function of $N_{\mathrm{train}}$, for the 6:6:6:6:1 architecture (using a single random seed).  The r.m.s.\ decreases as $N_{\mathrm{train}}$ increases but begins to level off for large $N_{\mathrm{train}}$ values.  Clearly there is a trade-off -- one can obtain fairly good results for quite small training sets (e.g.\ r.m.s.\ $\sim$0.15 for $N_{\mathrm{train}} = 1000$) but larger training sets can give significant improvements (e.g.\ r.m.s.\ $\sim$0.12 for $N_{\mathrm{train}} = 10000$).

Also plotted in Fig.\ \ref{fig.photz.anns_ntrain} are the r.m.s.\ values for different redshift bins: $z < 0.5$, $0.5 < z < 1.0$ and $z > 1.0$.  In the testing set, galaxies are distributed between these bins in the approximate ratio 2:2:1.  Since there are fewer training galaxies in the high-redshift bin, the network weights are less constrained and the r.m.s.\ values are larger.  From $N_{\mathrm{train}} = 200$ to $N_{\mathrm{train}} = 5000$ the r.m.s.\ in the $z < 0.5$ bin drops by about 20 per cent while the r.m.s.\ in the $z > 1.0$ bin drops by more than 50 per cent.  Thus larger training sets are important for tying down the redshifts of rare objects, but if this is not of particular interest (e.g.\ many large-scale structure surveys) then one can manage with smaller training sets.

It is important to note that the size of $N_{\mathrm{train}}$ required to achieve a given accuracy depends on the variation inherent in the data set.  In particular, for photometric redshifts it depends on the redshift range.  In $\S$\ref{sec.hyperz}---$\S$\ref{sec.sed}, in which the same $H < 20.5$ SAM catalogue is used, the training set size will be fixed to 10000.  For low redshift/shallow surveys (e.g.\ SDSS -- see $\S$\ref{sec.sdss}), smaller training sets should suffice.

\section[]{Comparisons with the template-\\fitting method} \label{sec.hyperz}
\begin{figure}
\vspace{5cm}
\includegraphics{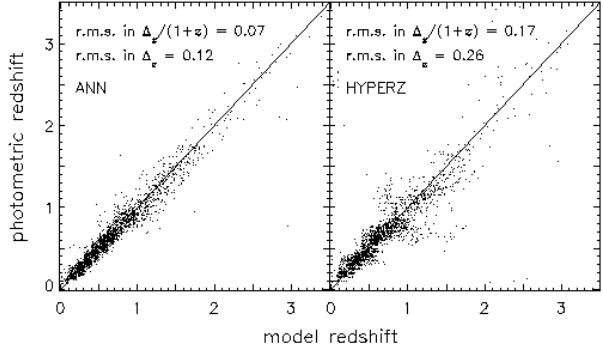}
\caption{\label{fig.photz.anns_bchyperz} A comparison of ANN and {\tt hyperz} photometric redshifts with model redshifts on 2000 test galaxies.  The simulated SAM catalogue is $H<20.5$ with 5$\sigma$ limits $U=25.1$, $B=26.6$, $V=26.1$, $R=25.6$, $I=24.7$ and $H=20.5$.  A committee of five 6:6:6:6:1 ANNs and a training set of 10000 galaxies were used. Eight Bruzual \& Charlot {\tt GISSEL'98} evolving SEDs were used as templates in {\tt hyperz} with $A_V$ in the range 0.0--1.2.}
\end{figure}

Fig.\ \ref{fig.photz.anns_bchyperz} compares the model redshifts of 2000 test galaxies with the redshifts estimated using a committee of five 6:6:6:6:1 ANNs and a training set of 10000 galaxies.  As found in $\S$\ref{sec.cf}, the r.m.s.\ in $\Delta_z = z_{\mathrm{model}} - z_{\mathrm{phot}}$ is 0.12 over the redshift range $0<z<3.5$.   Of course, the performance depends entirely on the particular set of filters and limiting magnitudes\footnotemark\footnotetext{The sample catalogue is of much poorer quality than the Hubble Deep Field, for example.}\ so, for comparison, the results of the template-fitting code {\tt hyperz\footnotemark} (Bolzonella, Miralles \& Pell\'o 2000)\footnotetext{\tt http://webast.ast.obs-mip.fr/hyperz/} are also shown (using the 8 synthetic Bruzual \& Charlot 1993 {\tt GISSEL'98} evolving templates, that are distributed with {\tt hyperz}, and dust extinction $A_V$ in the range 0.0--1.2).  The {\tt hyperz} results are comparable but slightly worse.  Formally the r.m.s.\ in $\Delta_z$ is 0.26, but this relatively high value may be due more to the small number of complete misidentifications -- against which the ANN seems to be more robust -- than to a general increased scatter.  However, as a further comparison, the scatter measured by the half-range of the central 68 per cent $\Delta_z$ values is 0.02 for the ANN and 0.11 for {\tt hyperz}.

\begin{figure}
\vspace{5cm}
\includegraphics{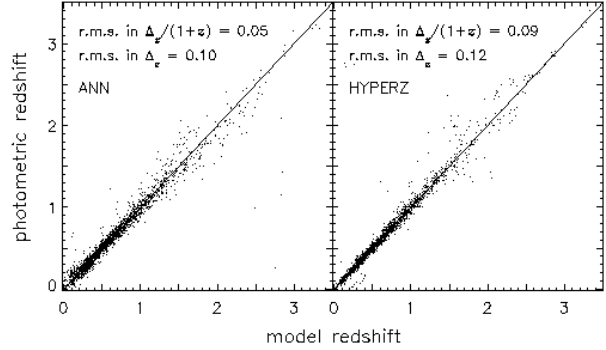}
\caption{\label{fig.photz.anns_cwwhyperz} A comparison of ANN and {\tt hyperz} photometric redshifts with model redshifts on 2000 test galaxies.  The simulated catalogue is $H<20.5$ with 5$\sigma$ limits $U=25.1$, $B=26.6$, $V=26.1$, $R=25.6$, $I=24.7$ and $H=20.5$.  Prior to adding photometric noise, the photometry for each galaxy in the semi-analytic model catalogue was replaced with the photometry of the best-fitting of four empirical CWW template spectra (E, Sbc, Scd or Im) at the same redshift.  A committee of five 6:6:6:6:1 ANNs and a training set of 10000 galaxies were used.  The 4 CWW SEDs themselves were used in {\tt hyperz}.}
\end{figure}

It may be argued that {\tt hyperz} is at a disadvantage here, relative to the ANN, since the ANN training and testing sets are based on the same spectral models while {\tt hyperz} is trying to fit a different set of templates.  For a real data set, {\tt hyperz} -- like all template-fitting methods -- could still suffer from template mismatches while ANNs automatically fit the data.  Synthetic templates were used in {\tt hyperz} for the above comparison since synthetic templates, albeit from a different source, are used in the semi-analytic model.  However when using {\tt hyperz} on real data sets, empirical SEDs, such as the four Coleman et al.\ (1980, CWW) SEDs -- E, Sbc, Scd and Im -- distributed with the {\tt hyperz} code, often produce better results.  It is possible that the four CWW templates are a closer match to real galaxy SEDs than the eight {\tt GISSEL'98} evolving templates are to the semi-analytic model SEDs.  Therefore as a further comparison, which maximally favours {\tt hyperz}, the $UBVRIH$ photometry for each galaxy in the semi-analytic model was replaced by $UBVRIH$ photometry for the best-fitting (using rest-frame $B - I$ colour) CWW template SED at the same model redshift.  Noise was added to the new photometry as described above and an $H<20.5$ sample was reselected.  The magnitudes and redshifts were normalized as described above and a new committee of five 6:6:6:6:1 ANNs was generated.  Fig.\ \ref{fig.photz.anns_cwwhyperz} compares the new ANN with {\tt hyperz}.  The formal r.m.s.\ in $\Delta_z$ is 0.10 for the ANN and 0.12 for {\tt hyperz}, while the half-width of the central 68 per cent $\Delta_z$ values is 0.02 for the ANN and 0.04 for {\tt hyperz}.  Hence, even for this case maximally favouring {\tt hyperz}, the ANN performs at least as well as template-fitting.

\section[]{Scatter due to photometric errors} \label{sec.hist}
\begin{figure}
\vspace{5cm}
\includegraphics{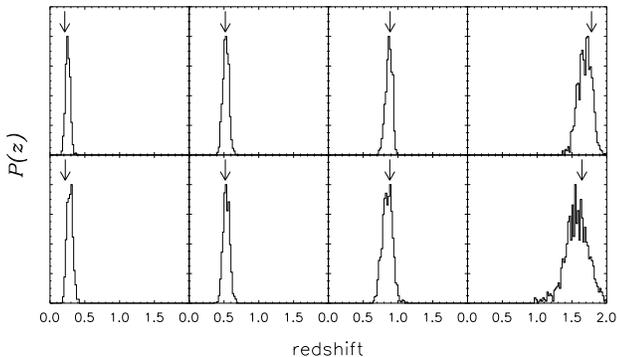}
\caption{\label{fig.photz.anns_mc} Eight galaxies were selected from the original `noiseless' SAM catalogue.  Random photometric noise was added simulating 5$\sigma$ limiting magnitudes $U=25.1$, $B=26.6$, $V=26.1$, $R=25.6$, $I=24.7$ and $H=20.5$, with 1000 simulations per initial galaxy.  The plots show histograms (one for each of the original eight galaxies) of the redshifts estimated with the committee of five ANNs previously trained on the full `noisy' $H < 20.5$ training set (as in $\S$\ref{sec.hyperz}).  The upper four panels are for bright ($H \sim 19.5$) galaxies while the lower four panels are for faint ($H \sim 20.5$) galaxies.  The arrows indicate the true (model) redshift of each galaxy.} 
\end{figure}

It is of interest to see what the distribution of estimated redshifts is for a particular galaxy, as a result of random photometric errors.  Random noise was added, as described above, to a selection of individual model galaxies, with 1000 random simulations per galaxy.  Then redshifts were estimated using the above previously trained committee of five 6:6:6:6:1 ANNs.  Fig.\ \ref{fig.photz.anns_mc} plots histograms of the estimated  redshifts for each original galaxy.  The width of the $P(z)$ distribution increases for fainter galaxies but a more pronounced trend is that it increases for galaxies at higher redshifts.  There are fewer training galaxies at high redshifts which means that (a) the network weights are less constrained than at lower redshifts, and (b) the training process gives more weight to improving the accuracy for the majority of galaxies at low redshifts at the expense of accuracy at high redshifts.

\section[]{Spectral type classification} \label{sec.sed}
One can also use ANNs to determine spectral type (independently of redshift) provided, as emphasized above, the training set is representative in terms of both spectral types {\it and} redshifts (and hence galaxy colours).  As in $\S$\ref{sec.hyperz}, the photometry for each semi-analytic model galaxy was replaced by that for the best-fitting CWW template SED at the same model redshift.  Then simulated photometric noise was added and an $H<20.5$ sample was selected.  The best-fitting CWW spectral types E, Sbc, Scd and Im give the desired output on which the network is trained.  

Since the input colour data is the same as in the photometric redshift problem of $\S$\ref{sec.cf}, a similar network architecture was used.  However it was modified to produce four outputs (viz.\ 6:6:6:6:4), corresponding to the four spectral types ({\it at any redshift})\footnotemark\footnotetext{Alternatively, since galaxy spectral types roughly follow a sequence (e.g.\ Connolly et al.\ 1995a; Naim et al.\ 1995), one could utilize a network with a single output node to classify galaxies, for example, on a scale of 0 to 1, where 0 corresponds to spectral type E and 1 corresponds to spectral type Im.}.  When training, the desired output is 1 for the output node corresponding to the correct type and 0 for the other three nodes.  When a galaxy of unknown spectral type is run through the ANN, the output in each node may be treated approximately as a probability for the galaxy being of the corresponding type, and the galaxy is assigned the type for which the probability is greatest (cf.\ e.g.\ Storrie-Lombardi et al.\ 1992; Lahav et al.\ 1996).  The input magnitudes were normalized to the range \mbox{[0, 1]} as above, 10000 training objects were used, and a committee of five 6:6:6:6:4 ANNs was generated.
 
The ANN results are displayed in Table \ref{table.photz.anns.sed}.  The ANN spectral types agree very well with the original CWW spectral types -- the mean error rate is $\sim$1 per cent.  Table \ref{table.photz.anns.sed} also shows the equivalent {\tt hyperz} results.  Here the ANN and {\tt hyperz} perform comparably well.

\begin{table*}
\caption{\label{table.photz.anns.sed} Comparisons of the efficiency with which ANNs and {\tt hyperz} recover galaxy spectral types. The simulated catalogue is $H<20.5$ with 5$\sigma$ limits $U=25.1$, $B=26.6$, $V=26.1$, $R=25.6$, $I=24.7$ and $H=20.5$.  A committee of five 6:6:6:6:4 ANNs and a training set of 10000 galaxies were used.}
\begin{center}
\begin{tabular}{llrrrrccrrrr}
&&\multicolumn{4}{c}{\tt hyperz}&&&\multicolumn{4}{c}{ANN}\\
\multicolumn{2}{l}{estimated SED}&E&Sbc&Scd&Im&&&E&Sbc&Scd&Im\\
\hline
true SED\\
E   && 428&   1&   0&   0&&& 422&   6&   0&   1\\
Sbc &&   9& 743&   1&   0&&&   3& 747&   3&   0\\
Scd &&   0&   3& 383&   0&&&   0&   1& 382&   3\\
Im  &&   0&   0&   9& 423&&&   3&   1&   7& 421\\
\hline
\end{tabular}
\end{center}
\end{table*}

\section[]{Performance of neural networks\\on SDSS data} \label{sec.sdss}
The Sloan Digital Sky Survey\footnotemark \footnotetext{Funding for the creation and distribution of the SDSS Archive has been provided by the Alfred P. Sloan Foundation, the Participating Institutions, the National Aeronautics and Space Administration, the National Science Foundation, the U.S. Department of Energy, the Japanese Monbukagakusho, and the Max Planck Society. The SDSS Web site is {\tt http://www.sdss.org/}. The Participating Institutions are The University of Chicago, Fermilab, the Institute for Advanced Study, the Japan Participation Group, The Johns Hopkins University, the Max-Planck-Institute for Astronomy (MPIA), the Max-Planck-Institute for Astrophysics (MPA), New Mexico State University, Princeton University, the United States Naval Observatory, and the University of Washington.} (SDSS; York et al.\ 2000) consortium have now publicly released more than 50000 spectroscopic redshifts along with $ugriz$ photometry and various image morphological parameters.  These provide an excellent opportunity to test ANNs on real data (see also Sowards-Emmerd et al.\ 2000 for a polynomial-fitting approach).  Objects were selected from the SDSS public data set using the following criteria: (1) the spectroscopic redshift confidence must be greater than 0.95 and there must be no warning flags, (2) $r < 17.5$, (3) redshift $< 0.5$.  Stars were left in with the galaxies but at these magnitudes they could have been fairly robustly removed using image morphology.  The order was randomized and the magnitudes, redshifts and other parameters were normalized to the range \mbox{[0, 1]}.  

Because the SDSS redshift range is much smaller than that of the SAM catalogue used in $\S$\ref{sec.cf} (though note the addition of stars here), a simpler architecture may suffice.  However, for the sake of simplicity, a similar architecture is used in this section also.  Two ANN architectures were used.  One -- 5:6:6:6:1 -- inputing $ugriz$ photometry, and the other -- 8:6:6:6:1 -- inputing $ugriz$ photometry and the SDSS pipeline star/galaxy classifier (`type') and Petrosian 50 per cent and 90 per cent $r$-band flux radii, $r_{50}$ and $r_{90}$.  A training set of size 10000 was used and, for each architecture, a committee of five ANNs was generated.

\begin{figure}
\vspace{5.cm}
\includegraphics{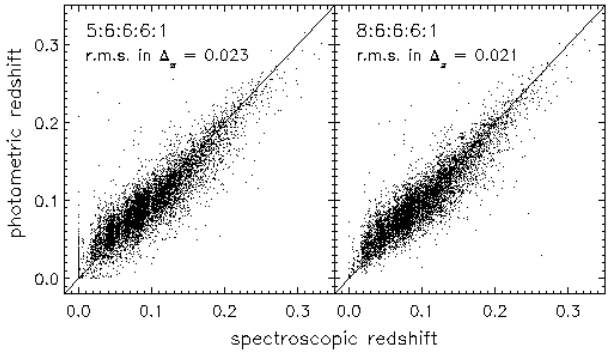}
\caption{\label{fig.photz.anns_sdss} A comparison of photometric and spectroscopic redshifts using SDSS public data.  Two ANN architectures were used, taking as input $ugriz$ photometry (5:6:6:6:1 architecture, left) and $ugriz$ photometry, SDSS star/galaxy classifier and Petrosian 50 and 90 per cent $r$-band flux radii (8:6:6:6:1 architecture, right).  A training set of size 10000 was used.  The ANNs were tested on a separate testing set of size 7000 (plotted).  In each panel, redshift estimates are medians from a committee of five ANNs.}
\end{figure}

Fig.\ \ref{fig.photz.anns_sdss} compares the ANN redshifts with spectroscopic redshifts for a testing set of 7000 galaxies.  The r.m.s.\ in $\Delta_z$ are 0.023 and 0.021 for committees of five 5:6:6:6:1 and 8:6:6:6:1 networks respectively, while the mean offsets are both 0.000.  These results are easily as good as, and probably better than, the best results that template-fitting photometric redshift methods can produce.  There are also very few outliers.  The spike at $z=0$ in the 5:6:6:6:1 network results is due to misidentified stars.  In the 8:6:6:6:1 network the addition of morphological parameters largely removes this feature.

While 10000 training objects were used for the above networks, one can still do fairly well with much smaller training set sizes.  For example, with only 500 training objects, the r.m.s.\ values become respectively 0.028 and 0.027 (cf.\ Fig. \ref{fig.photz.anns_ntrain}).  Again, larger training sets are expected to be useful for pinning down the classifications of rare objects -- e.g.\ the r.m.s.\ in the redshift range $0.25 < z < 0.35$ is $\sim$0.03 for $N_{\mathrm{train}} = 10000$ but degrades to $\sim$0.06 for $N_{\mathrm{train}} = 500$.

\section{Conclusions} \label{sec.conclude}
ANNs can produce photometric redshift accuracies that are comparable to or better than template-fitting procedures.  However they do rely on large and representative training samples and an ANN is only applicable to the particular survey filters and redshift range upon which it has been trained.  For large photometric/spectroscopic surveys, such as the SDSS and future deeper surveys such as DEEP2\footnotemark\footnotetext{\tt http://astron.berkeley.edu/$\sim$marc/deep/} and the VIRMOS-VLT Deep Survey\footnotemark\footnotetext{\tt http://www.astrsp-mrs.fr/virmos/}\ (VVDS; Le F\`evre et al.\ 2000), where large spectroscopic samples are available, it seems that ANNs offer some significant advantages over previous approaches.  The VVDS, for example, is expected to include $> 100000$ redshifts to $I_{AB} = 22.5$, $> 40000$ redshifts to $I_{AB} = 24$ and $> 1000$ redshifts to $I_{AB} = 25$, providing ample training set sizes for the complementary deep imaging in $UBVRIK_s$ to $I_{AB} = 25$ (similar to the limits used in $\S$\ref{sec.sam}).

With careful modelling of photometric errors and some loss in the Bayesian statistics, bright spectroscopic samples may also be extrapolated to provide training sets for fainter photometric samples.

In addition, ANNs may also be used where spectroscopic redshifts are unavailable, by utilizing a simulated catalogue (e.g.\ semi-analytic model) as a training set.  By using theoretical SEDs in the training set, this method has all the disadvantages and advantages of standard template-fitting, but it also has the extra advantages (i) the `template' SEDs include a (more or less) realistic distribution of {\it complex} star formation histories, dust modelling and metallicities etc., giving fully Bayesian statistics, and (ii) the weights applied to different filters (and non-linear combinations thereof) may be more optimal than simple $\chi^2$-weighting.

To conclude, while template-fitting photometric redshifts may be used to good effect in pioneering studies of new populations of objects, spectroscopic confirmation will always be necessary to obtain truly robust scientific results.  Instead the real power of photometric redshifts lies in extending small very resource-intensive faint spectroscopic surveys to much larger fields-of-view and sample sizes.  That is to say, the area where photometric redshifts can best be used for robust and useful scientific gains is in the training-set regime.  ANNs provide a powerful tool for obtaining high-quality photometric redshifts in such surveys.

\section*{Acknowledgments}
We are grateful to B.\ D.\ Ripley for providing the ANN training code, Avi Naim for related programmes and advice, Richard Ellis and Richard McMahon for discussions, and the anonymous referee for many useful suggestions.  AF was supported by Trinity College, Cambridge, an Isaac Newton Studentship, and an L.\ B.\ Wood Travelling Scholarship during this work, and also gratefully acknowledges the use of facilities at the University of Otago.

\bsp

\label{lastpage}

\end{document}